# CALLAS OU LA TRAJECTOIRE DU METEORE

Jean-Philippe EPRON[1], Jocelyne SARFATI[2], Nathalie HENRICH[3]

(1) ORL PHONIATRE Le parc Ravel 6 chemin du PILLON 74200 THONON
(2) ORL PHONIATRE 10 rue de Belgrade 38000 GRENOBLE
(3) Chargée de Recherche CNRS, Département Parole et Cognition, GIPSA-lab, 961 rue de la Houille Blanche, Domaine universitaire, BP 46, 38402 SAINT MARTIN D'HERES

**Résumé**

La carrière lyrique de Maria Callas, aussi exceptionnelle qu'elle ait été, a été également remarquable par sa brièveté. Les premiers signes de déclins sont apparus dès l'âge de 36 ans et la voix se tait à 40 ans seulement. Si la littérature a abondamment commenté cette détérioration précoce, elle a peu analysé ses caractéristiques. Le but de notre étude est de réaliser une analyse perceptive puis acoustique d'airs enregistrés par l'artiste à son zénith, puis lors du déclin. Nous avons établi une analyse descriptive de différentes altérations perçues à l'écoute, puis confronté ces observations aux analyses acoustiques.

**Mots-clés :** *Callas, déclin, analyse acoustique, analyse spectrographique*

**Abstract**

The lyric career of Maria Callas, though exceptional, is also noteworthy for its brevity. The first signs of downturn appeared at the age of 36 and her voice fell silent at only 40. Though the literature has massively commented on this premature worsening, few analyses of its characteristics have been made public so far. The purpose of our study was to realise a perceptual and acoustical analysis of recorded arias by the artist at the climax to the fall. The audible impairments were first verbally described, and then compared to acoustical observations based on spectrographic analyses and fundamental-frequency measurements.

## INTRODUCTION

La carrière de Maria Callas a été une des plus brèves du théâtre lyrique au XXème siècle. Elle fait ses débuts à Athènes en 1942 dans le rôle de Tosca, âgée seulement de 18 ans. Sa carrière prend ensuite son envol en Italie en 1947 avec Gioconda, Aida et surtout I Puritani en 1948, qui vont révéler sa vraie nature de chanteuse belcantiste. Elle connaîtra son zénith au milieu des années cinquante, sa maturité artistique et interprétative coïncidant avec des moyens préservés.
Toutefois, dès 1956, on note une fatigue vocale plus fréquente et une certaine diminution des moyens vocaux.

A l'orée des années 60, il est indiscutable que la voix est nettement altérée. Les apparitions sur scène sont beaucoup plus rares, laissant une large place au récital. Seuls quelques rôles sont conservés à son répertoire. Elle est âgée seulement de 36 ans.

Ses dernières apparitions dans un rôle intégral auront lieu en 1965, à l'âge de 41 ans.

Elle n'apparaîtra plus sur scène que pour une série de récitals de 1973 à 1974, dans lesquels la voix n'est plus qu'un lointain écho de ses anciennes splendeurs. Elle a 50 ans.

Les critiques de ses enregistrements et de représentations qu'elle a données font état de ce déclin vocal précoce. Toutefois, la description en est souvent peu précise, lyrique ou poétique et assez rarement analytique.

Force nous est de reconnaître que les compte-rendus de ses derniers disques s'attachent plus à la description de ses qualités interprétatives qu'à ses qualités vocales proprement dites.

Nous avons voulu porter un regard plus analytique sur différents paramètres de ce déclin vocal et livrer quelques observations et réflexions que nous avons pu faire à ce sujet.

## MATERIEL ET METHODES

Nous avons dans un premier temps procédé à une écoute par un jury composé de deux amateurs d'opéra des mêmes airs ou passages musicaux enregistrés par la chanteuse au cours des années 50 puis au milieu des années 60.

Les oeuvres retenues étaient les suivantes :

- Puccini (Tosca intégrale), enregistrement studio 1953 EMI, enregistrement studio 1964 EMI

- Verdi (Nabucco), Récitatif et Air de l'acte II : « Ben io t'invenni », enregistrement live 1952, enregistrement studio 1958 EMI, enregistrement live 1963

- Verdi (Un ballo in maschera), Air de l'acte II : « Ecco l'orrido campo »,enregistrement studio 1956 EMI, enregistrement studio 1964 EMI

- Verdi (Il trovatore), Air de l'acte IV : « Vanne lasciami », rnregistrement live Scala Milan 1953, enregistrement studio 1956 EMI, enregistrement studio 1964 EMI

Nous avons porté notre attention sur les paramètres suivants : l'intonation, la précision des attaques, le legato, le vibrato, le souffle essentiellement sur les enregistrement live (le studio permettant lors du montage des collages qui masquent les reprises respiratoires).

Nous émettons des réserves sur l'analyse du timbre qui peut être flatté ou au contraire desservi par la qualité originelle de l'enregistrement.

Nous ne nous sommes pas attachés aux paramètres de puissance et de projection vocale. Même si leur diminution est attestée par l'ensemble des contemporains de Callas, ce caractère est bien trop dépendant de la prise de son de l'enregistrement.

Dans un deuxième temps, nous avons procédé à une analyse acoustique de ces enregistrements. Nous avons pour cela sélectionné des fragments musicaux sans accompagnement orchestral et réalisé une analyse spectrographique (décomposition temps-fréquence du signal) et des mesures acoustiques. Le logiciel WaveSurfer a été utilisé pour une analyse visuelle et qualitative des spectrogrammes, et le langage de programmation Matlab pour une étude quantitative des variations de fréquence fondamentale.

**RESULTATS**

**Analyse perceptive**

**L'intonation** : Elle est régulièrement prise en défaut dans le registre aigu. Nous avons la perception d'une intonation souvent trop basse : Nous avons relevé les défauts d'intonation les plus flagrants dans le Tableau 1.

**Les attaques, le legato** : Les attaques des notes forte sont souvent dures avec parfois des coups de glotte. Ces problèmes sont déjà perceptibles en 1958 dans l'extrait de Nabucco. Elle n'a pourtant pas encore « perdu sa voix » … Le Tableau 2 résume les anomalies les plus notables. On note aussi fréquemment une imprécision dans l'attaque de la note, un peu en dessous. Cela se traduit dans certaines vocalises par un manque de netteté de la vocalise, qui apparaît « savonnée ». Le legato est parfois rompu dans certaines phrases du fait de la dureté des attaques.

**Le vibrato** : Il apparaît souvent élargi, ralenti et irrégulier, en particulier sur les notes *forte*. (TOSCA : duos avec Cavaradossi , et Scarpia, NABUCCO : vocalise ascendante jusqu'au contre-ut)

**Le souffle** semble plus court : La vocalise ascendante de Nabucco est coupée. Dans le Vissi d'arte de TOSCA, la phrase « perche signore » est interrompue lors de l'enregistrement de 1964, pourtant en studio, ce qui n'était pas le cas en 1953.

**Le timbre** semble appauvri, moins riche. On note la présence du fameux son « dans les joues » dans le medium qui caractérise sa dernière période : les voyelles [a] et [o] ouvert sont souvent très fermées et transformées en un son unique proche du [o] fermé et du [oe].

*Tableau 1 : défauts d'intonation*

| TOSCA acte II | **Ah** piuttosto giu mi avvento | Si bémol |
|---|---|---|
| TOSCA acte III | Io quelle **la**ma gli piantai nel cor | Contre-ut |
| TOSCA acte III | Fur **va**ni scongiuri e pianti | La bemol |
| TROVATORE | Récitatif pie**to**so | Si bémol |
| TROVATORE | Aria **con**forto | La bémol |
| TROVATORE | Aria ma deh non dirgli improvi**do** | Si bémol et contre-ut |
| BALLO IN MASCHERA | Aria terribil **sta** | si |
| BALLO IN MASCHERA | Aria miserere d'un **po**vero cor **ah** signor | Contre-ut |
| NABUCCO | Récitatif iniqui **tu**tti | Si bémol |
| NABUCCO | Récitatif sdegno fa**tal** | Contre-ut |

*Tableau 2 : attaques dures*

| NABUCCO 1958 | Récitatif **A** sterminar Giudea | sol |
|---|---|---|
| NABUCCO 1958 | Récitatif Che son io **qui** | la |
| NABUCCO 1963 | Récitatif **sde**gno fatal | mi |
| TOSCA 1964 acte I | Duo e troppo **be**lla | sol |
| TOSCA 1964 acte I | Duo e l'atta**van**ti | Si bémol |
| TOSCA 1964 acte II | Duo **So**lo si | Si bémol |
| TOSCA 1964 acte III | Duo fur **va**ni | La bémol |
| TOSCA 1964 acte III | Duo sei **mio** | La bémol |
| BALLO IN MASCHERA | Récitatif s'a**dem**pia | La bémol |
| BALLO IN MASCHERA | Récitatif **che ti** resta | La bémol |
| BALLO IN MASCHERA | Récitatif **T'a**nnienta | La bémol |

**Analyse acoustique**

L'analyse acoustique a été axée sur l'examen de 3 paramètres principaux : l'intonation, le vibrato, et l'attaque du son.

- **L'intonation :**

Les analyses acoustiques comparatives montrent une baisse de précision dans la justesse des notes brèves et tenues, et dans leurs attaques. Dans l'extrait sélectionné de Nabucco, 1963, par exemple, le Contre-ut est attaqué en dessous (voir Fig. 1). Maria Callas atteint ensuite la note juste, mais elle ne la maintient que brièvement, sur deux cycles de vibrato et baisse en fin de tenu. Les notes brèves sont parfois chantées trop basses, ainsi que l'illustre la vocalise descendante extraite d'Un Ballo in maschera en 1964 (voir Fig.2). Des notes parfois trop hautes sont également mises en évidence par les analyses spectrographiques, comme pour le Si aigu dans Tosca, 1964, par exemple.

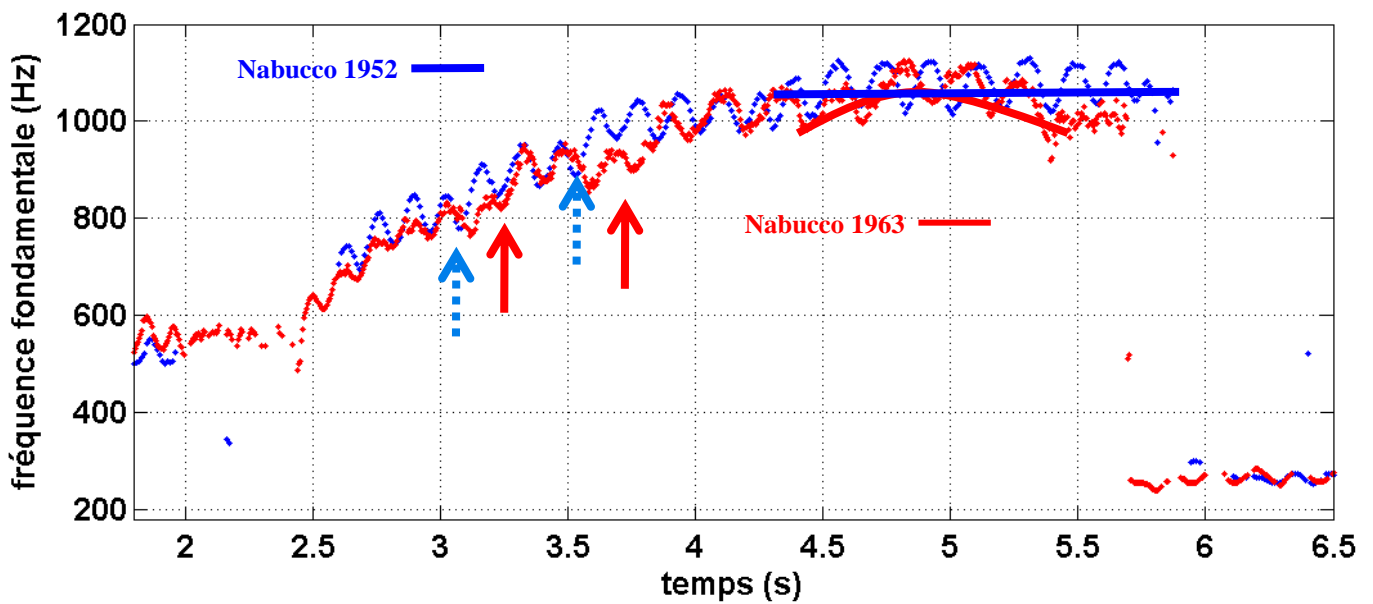

*Figure n°1 : Mesure de la fréquence fondamentale sur deux extraits sans accompagnement instrumental de Nabucco, enregistrés en 1952 et 1963. Les lignes présentent la stabilité ou instabilité du vibrato sur la note finale (contre-ut). Les flèches indiquent la gestion du vibrato (continuité ou rupture de rythme).*

- **Le vibrato :**

Il apparaît irrégulier voire totalement désorganisé dans les extraits issus de la période de déclin vocal (voir par exemple l'extrait de Nabucco, 1963 sur la Fig.1). Il est également souvent ralenti, ainsi que l'illustre la Fig.2 dans le cas d'un extrait d'Un Ballo in maschera en 1964. Parfois, il est totalement absent, comme dans le Si aigu de l'extrait de Tosca, 1964, présenté sur la Fig.3. Sur les extraits sélectionnés, nous n'avons pas noté d'augmentation de son amplitude.

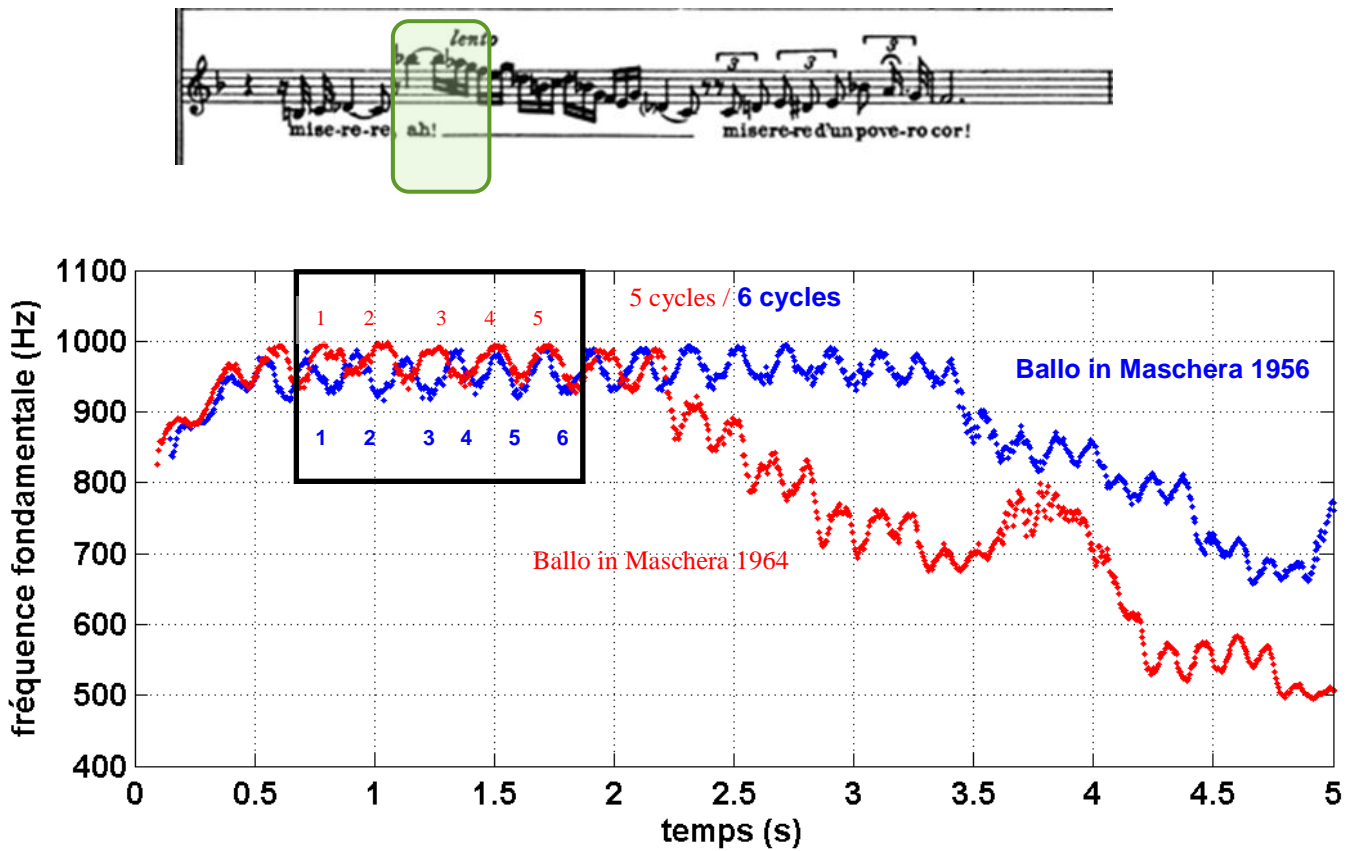

*Figure n°2 : Mesure de la fréquence fondamentale sur deux extraits sans accompagnement instrumental d'Un ballo in maschera, enregistrés en 1956 et 1964. Les cycles de vibrato sont numérotés dans le rectangle de sélection, pour illustrer la différence de fréquence du vibrato entre les deux enregistrements.*

- **L'attaque du son :**

L'attaque des certaines notes est imprécise, ainsi que l'illustre l'analyse de la vocalise ascendante dans Nabucco, 1963, présentée sur la Fig.1. On observe une désynchronisation entre le vibrato et l'attaque de la note. Dans la vocalise ascendante extraite de Nabucco, 1952, le passage à la note supérieure se fait systématiquement sur la phase ascendante du vibrato, donc en phase avec ces vibrations. En 1963, le passage se fait sur la phase descendante du vibrato, en opposition de phase avec le vibrato, et entraînant ainsi une interruption du cycle vibratoire. Sur l'exemple extrait de Tosca, 1964, les transitoires sont moins précis, et les attaques se font par en-dessous (voir Fig.4). On note un rallongement des temps de montée à la note.

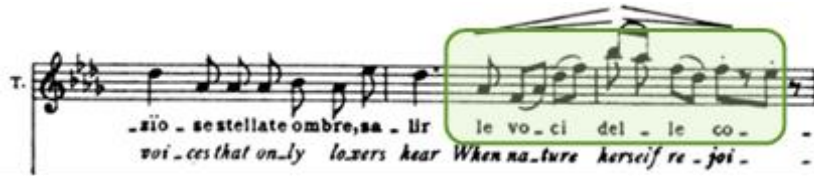

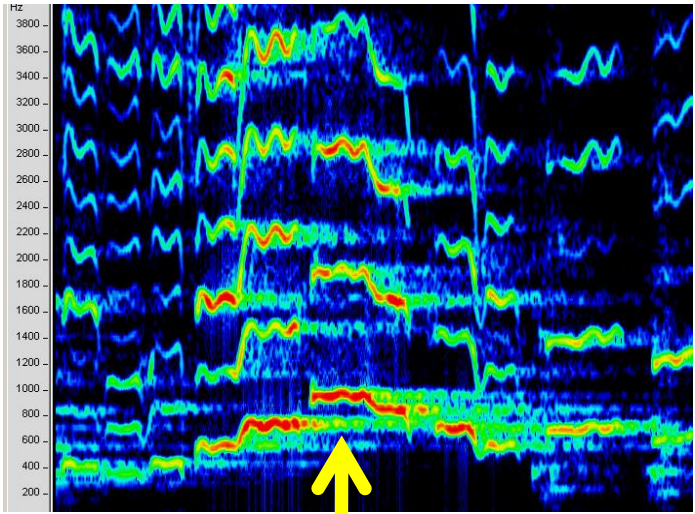
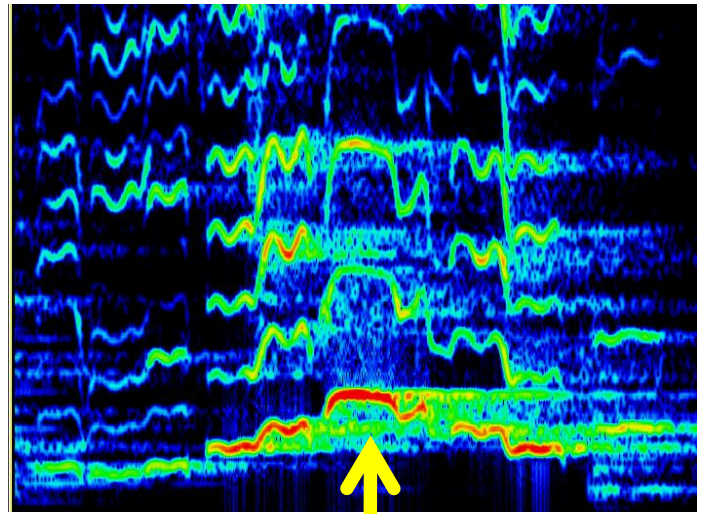

**Tosca 1953**  **Tosca 1964**

*Figure n°3 : Analyse spectrographique comparative du même extrait de Tosca (Acte 1, « le voci delle co- », enregistré en 1953 et 1964.*

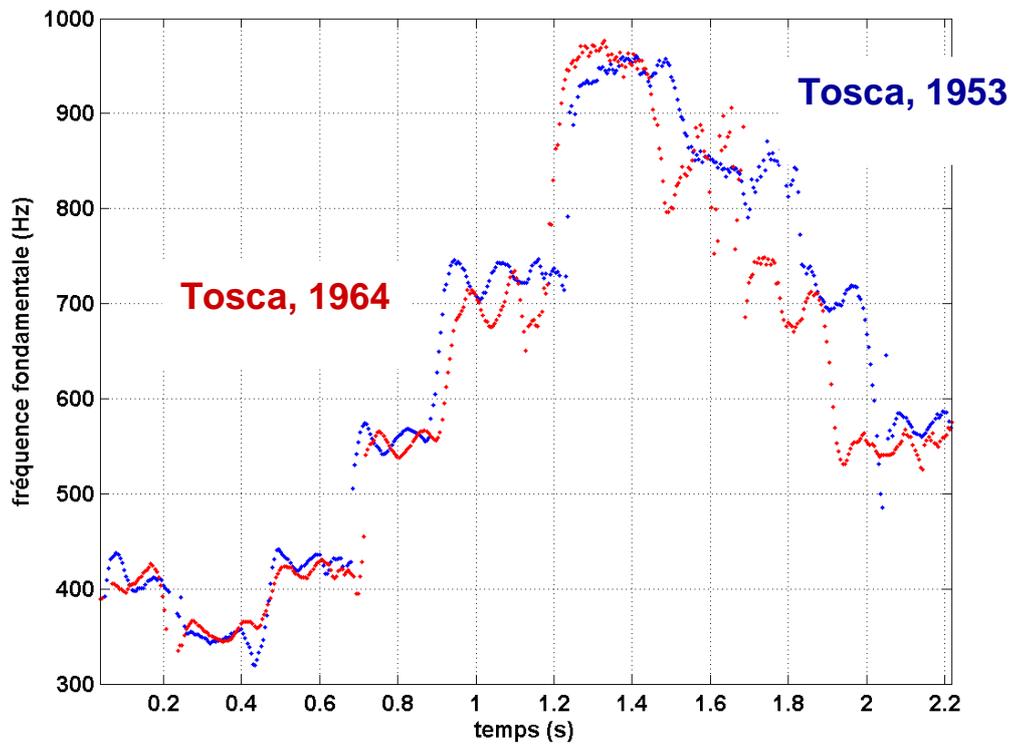

*Figure n°4 : Mesure de la fréquence fondamentale sur deux extraits sans accompagnement instrumental de Tosca (Acte 1, « le voci delle co- », enregistré en 1953 et 1964.*

**DISCUSSION**

La littérature consacrée à Callas est extrêmement abondante. Elle fait état de ce déclin vocal précoce, déjà perceptible au milieu des années 50.
On parle en effet du grinçant des aigus, d'une perte de volume et de sécurité [3].
Lorcey [2] souligne la pauvreté de la matière sonore, la sensation de gosier serré, les hurlements et les stridences dans les aigus. Rostand [3] fait état d'aigus blancs et plats.
Le vibrato est accentué, élargi, « propre à donner le mal de mer » dans les mauvais soirs [3].
On parle aussi de voix qui bouge [2], de sons indéfinissables quant à la hauteur [3].
Mancini [3] fait remarquer à propos de ses ultimes Norma au palais Garnier que « les pianissimi impalpables remplissaient la scène, alors que les sons forcés ou appuyés sur un souffle déficient se perdaient dans la salle ».
J. Ardouin [1], pourtant toujours prêt à voler au secours de la diva, écrit à propos des ultimes récitals, que sa voix est fantomatique, instable et manque de soutien. Sa respiration est erratique et la plupart des notes aiguës sont réduites à des cris.

Notre analyse nous a permis de préciser les notions suivantes : le caractère le plus fréquemment rencontré est l'imprécision de l'intonation dans le registre aigu (sur le plan strictement fréquentiel, de nombreuses notes sont trop basses). Nous avons également observé dans un extrait une note trop haute. Sur tous les exemples comparés, la chanteuse éprouve une difficulté à stabiliser une note tenue à une hauteur donnée. Le vibrato est généralement ralenti et moins régulier, ce qui participe à l'instabilité de la voix et à cette sensation de « voix qui bouge ». Si son élargissement n'est pas mis en évidence dans l'analyse acoustique des extraits sélectionnés, il est nettement perçu à l'écoute dans les notes *forte*. Il participe probablement au défaut d'intonation. Les attaques sont durcies, explosives, avec parfois des coups de glotte. Elles deviennent imprécises, souvent trop basses en fréquence et la note est lors ajustée secondairement. La consonne est allongée, ne laissant pas s'épanouir librement la note et rompant parfois le legato dans des passages plus dramatiques. La question qui s'impose toujours en constatant ce déclin vocal prématuré est bien sûr : pourquoi ?
Mais il est à craindre qu'elle demeure sans réponse. On ne peut qu'échafauder des hypothèses sur les origines de cette dégradation vocale. La littérature a souvent mis en cause la perte progressive du soutien et le célèbre régime : la perte de poids extrêmement rapide, semble t il obtenue par une association d'hormones thyroïdiennes, de diurétiques et d'amphétamine aurait pu entraîner une fonte musculaire et une altération de la conduction neuromusculaire

Néanmoins, sans tenir compte de ces suppositions physiologiques, le comportement de forçage clairement audible dans les derniers enregistrements n'est il pas aussi le défaut de ses qualités ? L'esthétique de Callas privilégiait l'expression musicale et théâtrale au détriment du beau son pur. Elle sculptait la ligne musicale en orfèvre et attachait un grand soin à l'articulation des consonnes, parfois explosives, préservant l'intelligibilité du texte et lui donnant son expression musicale. L'accentuation progressive de ce phénomène a pu entraîner un durcissement des attaques, parfois présent lors d'enregistrements précoces, notamment véristes (Mascagni : Cavalleria rusticana, enregistrement studio 1953).

**CONCLUSION**

Jacques Bourgeois, critique musical et ami de la diva, déclarait à propos de ses Norma au Palais Garnier en 1964 : « *Maria Callas est comme l'acropole : elle est encore plus belle quand elle est en ruines.* »

A l'écoute de ses enregistrements du milieu des années 60, parfois cruels lorsqu'on se réfère aux splendeurs vocales qu'elle laissait en témoignage 10 ans plus tôt, nous sommes encore bouleversés par la beauté et l'intelligence de son interprétation. Le soin apporté à la ligne de chant résiste aux défauts d'intonation et à un vibrato envahissant. Il témoigne d'une grande musicienne et d'une grande technicienne en lutte avec un instrument qui se dérobe. La cause en reste encore mystérieuse, et la chanteuse a emporté son secret dans la tombe. Mais nous pouvons encore la saluer et lancer un dernier « Brava Callas ».

**BIBLIOGRAPHIE**